\documentstyle[emulateapj,apjfonts,psfig]{article}
\def\deg{\ifmmode^{\circ}\;\else$^{\circ}\;$\fi} 

\righthead{FORMATION OF MILLISECOND PULSARS WITH HEAVY WHITE DWARF COMPANIONS}
\lefthead{TAURIS, VAN DEN HEUVEL \& SAVONIJE}

\submitted{Accepted by {\rm The Astrophysical Journal Letters}, December 20, 1999}

\begin{document}

\title{Formation of Millisecond Pulsars with Heavy White Dwarf Companions.\\
       Extreme Mass Transfer on Sub-Thermal Timescales}

\author{Thomas M. Tauris\altaffilmark{1},
        Edward P. J. van den Heuvel\altaffilmark{1},
        Gerrit J. Savonije\altaffilmark{1}
        }

\altaffiltext{1}{Center for High-Energy Astrophysics,
and Astronomical Institute ``Anton Pannekoek'',
University of Amsterdam,
Kruislaan 403, NL-1098 SJ Amsterdam, The Netherlands;
tauris@astro.uva.nl, edvdh@astro.uva.nl, gertjan@astro.uva.nl}

\begin{abstract}
We have performed detailed numerical calculations of the non-conservative 
evolution of close {X}-ray binary systems with intermediate-mass 
(2.0 -- 6.0$\,M_{\odot}$) donor stars and a $1.3\,M_{\odot}$ accreting
neutron star.
We calculated the thermal response of the donor star to mass loss,
in order to determine its stability and follow the evolution of
the mass transfer.
Under the assumption of the ``isotropic re-emission model'' we 
demonstrate that in many cases it is possible for the binary to
prevent a spiral-in and survive a highly super-Eddington mass-transfer phase 
($1\ll \dot{M}/\dot{M}_{\rm Edd} < 10^5$) on a sub-thermal timescale,
if the convective envelope of the donor star is not too deep.
These systems thus provide a new formation channel for binary
millisecond pulsars with heavy {CO} white dwarfs and relatively short
orbital periods (3--50 days). However, we conclude that to produce 
a binary pulsar with a {O-Ne-Mg} white dwarf or $P_{\rm orb}\sim 1$ day
(e.g. PSR B0655+64) the above scenario does not work, and a spiral-in phase 
is still considered the most plausible scenario for the formation of such a system.

\end{abstract}

\section{Introduction}
Recently a large number of binary millisecond pulsars
(BMSPs) with relatively heavy white dwarf (WD)
companions have been reported. These pulsars form a distinct class
of BMSPs (cf. Table~1) which are characterized by relatively slow spin periods
($P_{\rm spin}\simeq 10-200$ ms) and high period derivatives:
$10^{-20} < \dot{P}_{\rm spin} < 10^{-18}$.
It has been suggested that such BMSPs evolved through a 
common envelope and spiral-in phase (e.g. van~den~Heuvel 1994).
This gives a natural explanation for their close orbits and the
presence of a relatively heavy {CO/O-Ne-Mg} WD, if the mass transfer
was initiated while the donor star (the progenitor of the WD)
ascended the AGB.
However, it has been argued that a neutron star engulfed in a common envelope
might experience hypercritical accretion and thereby collapse into a black hole
(e.g. Chevalier 1993; Brown 1995). If this picture is correct then these
BMSPs can not have formed in a common envelope and spiral-in phase. 

Here we investigate an alternative scenario for producing the mildly 
recycled BMSPs with
{He} or {CO} WD in close orbits, in which a 2--6$\,M_{\odot}$
donor star, with a non- (or partly) convective envelope, transfers mass on a 
sub-thermal timescale and yet in a dynamically stable mode.

\section{Stability Criteria and Mode of Mass Transfer}
The stability and nature of the mass transfer is very important
in binary stellar evolution. It depends on the response of the
mass-losing donor star and of the Roche-lobe (e.g.
Paczynski 1976; Soberman, Phinney \& van~den~Heuvel 1997). 
The mass transfer is stable as long as the
donor star's Roche-lobe continues to enclose the star. Otherwise it
is unstable and proceeds on the shortest unstable timescale.

As long as the mass of the donor,
$M_2$ is less than $1.8\,M_{\odot}$ the mass transfer will be dynamically
stable for all initial orbital periods (e.g. Tauris \& Savonije 1999). 
These LMXBs are the progenitors of the BMSPs with a helium WD 
companion.
The observational absence of {X}-ray binaries with Roche-lobe filling
companions more massive than $\sim 2\,M_{\odot}$ has been attributed
to their inability to transfer mass in such a stable mode
that the system becomes a persistent long-lived X-ray source
(van~den~Heuvel 1975; Kalogera \& Webbink 1996). 
Below we investigate for these systems how the stability of the
RLO depends on the evolutionary status of
the donor (and hence the orbital period) at the onset of mass transfer. 

\subsection{Numerical computations}
We have calculated the evolution of a large number of {X}-ray binaries
with a donor star of mass $2\le M_2/M_{\odot} < 6$ and 
a $1.3\,M_{\odot}$ accreting neutron star.
Both the radius of the donor star as well as its Roche-lobe
are functions of time and mass (as a consequence of nuclear burning, 
magnetic braking and other tidal spin-orbit couplings).
We used an updated version of Eggleton's numerical computer code
(Pols~et~al. 1998) to keep track of the stellar evolution and
included a number of binary interactions to carefully follow the
details of the mass-transfer process. 
For all donor stars considered here we assumed a chemical composition
of {X}=0.70 and {Z}=0.02, and a mixing-length parameter of 
$\alpha =l/H_{\rm p}=2.0$.
We refer to Tauris \& Savonije (1999)
for a detailed description of our computer code.

\subsection{Highly super-Eddington mass transfer}
The maximum accretion rate onto a neutron star
is given approximately by the Eddington limit for
spherical accretion of a hydrogen gas,
$\dot{M}_{\rm Edd}=1.5\times 10^{-8}\,M_{\odot}$ yr$^{-1}$. 
If the mass-transfer rate from the donor star, $\dot{M}_2$ is larger than 
this limit, radiation pressure from the accreted material will cause
the infalling matter to be ejected from the system at a rate:

$|\dot{M}|=|\dot{M}_2|-\dot{M}_{\rm Edd} \simeq |\dot{M}_2| 
   \mbox{\hspace{0.5cm}if $\dot{M}_2 \gg \dot{M}_{\rm Edd}$}$ \\
In systems with very large mass-transfer rates, matter piles up around the 
neutron star and presumably forms a growing, bloated cloud engulfing a large
fraction of the accretion disk.
A system will only avoid a spiral-in if it manages to evaporate
the bulk of the transferred matter via the liberated accretion energy.
This would require the radius of the accretion cloud,
$r_{\rm cl} > R_{\rm NS}\,|\dot{M}_2|/\dot{M}_{\rm Edd}$
in order for the liberated accretion energy to eject the transfered
material
($\sim 0.1\,\dot{M}_{\rm Edd}\,c^2 \ga \frac{1}{2}\,\dot{M}_2\,v_{\rm esc}^2$,
where $v_{\rm esc}^2=2GM_{\rm NS}/r_{\rm cl}$; $R_{\rm NS}$ is the
radius of the neutron star).
If the material which is to be ejected comes closer to the neutron star
it will have too much negative binding energy in order for the
liberated accretion energy to expel it.
At the same time $r_{\rm cl}$ must be smaller than the Roche-lobe
radius of the neutron star during the entire evolution, if formation
of a common envelope (CE) is to be avoided\footnote{
Note, that if no efficient cooling processes are present in
the accretion disk then the incoming matter retains its net (positive)
energy and is easily ejected in the form of a wind from the disk
(Narayan \& Yi 1995; Blandford \& Begelman 1999).
Even if the arriving gas is able to cool, interactions between
the released radiation from this process and the infalling gas
may also help to eject the matter. In both cases $r_{\rm cl}$ can be
smaller than estimated above.}.\\
A simple isotropic re-emission model will approximately
remain valid for our scenario. 
In this model it is assumed that matter flows over 
conservatively from the donor star to the vicinity of the neutron star
before it is ejected with the specific orbital
angular momentum of the neutron star. 
Assuming this to be the case we find that, even for
extremely high mass-transfer rates ($|\dot{M}_2| > 10^4\,\dot{M}_{\rm Edd}$),
the system can avoid a CE and spiral-in evolution.

\section{Results}
\subsection{A case study: $M_2=4.0\,M_{\odot}$ and $P_{\rm orb}=4.0$ days}
In Fig.~1 we show the
evolution of a binary initially consisting of a neutron star and
a zero age main-sequence companion star with masses $M_{\rm NS}=1.3\,M_{\odot}$
and $M_2=4.0\,M_{\odot}$, respectively, and initial orbital period
$P_{\rm orb}=4.0$ days.

At the age of $t=176.6$ Myr the companion has evolved to fill its Roche-lobe
($R_2=8.95\,R_{\odot}$; $T_{\rm eff}=9550$ K) 
and rapid mass transfer is initiated (A).
The donor star has just evolved past the MS hook in the HR-diagram
and is burning hydrogen in a shell around a $0.56\,M_{\odot}$ helium core.
Prior to the mass-transfer phase, a radiation-driven wind 
($|\dot{M}_2|\sim 4\times 10^{-10}\;M_{\odot}$ yr$^{-1}$) has caused the donor
to decrease its mass slightly ($M_2=3.99\,M_{\odot}$) and consequently 
resulted in a slight widening of the orbit ($P_{\rm orb}=4.02$ days).
Once the donor fills its Roche-lobe it is seen to lose mass at a very high rate
of $|\dot{M}_2|\simeq 4\times 10^{-5}\;M_{\odot}$ yr$^{-1} = 2.7\times 10^3\,
\dot{M}_{\rm Edd}$. At this stage
the donor has only developed a very thin convective envelope of size 
$Z_{\rm conv}=0.015\,R_{\odot}$,
so its envelope is still radiative and will therefore shrink in 
response to mass loss.
At $t=176.7$ Myr its radius has decreased to a minimum value of
$3.38\,R_{\odot}$, but now $Z_{\rm conv}=0.97\,R_{\odot}$.
At this point (s) the donor has a mass of $1.76\,M_{\odot}$,
$T_{\rm eff}=5640$ K, and $P_{\rm orb}=1.59$ days.
The donor expands again, but shortly thereafter, its rate of
expansion slows down causing 
$|\dot{M}_2|$ to decrease to $\sim 10\,\dot{M}_{\rm Edd}$.

The mass transfer ceases (B) when the donor has an age of 178.3 Myr.
At this stage $P_{\rm orb}=8.11$ days, $R=6.68\,R_{\odot}$,
$Z_{\rm conv}=0.07\,R_{\odot}$ and $T_{\rm eff}=12700$ K. 
The mass of the donor is
$0.618\,M_{\odot}$. It still has a $0.56\,M_{\odot}$ helium core,
but now only a $0.06\,M_{\odot}$ envelope consisting of
16\% {H} and 82\% {He}.
The mass-transfer phase (A--B) lasts relatively short:
$t_{\rm X}=1.7$ Myr, and hence the neutron star will only accrete:
$\Delta M_{\rm NS} = t_{\rm X}\;\dot{M}_{\rm Edd} = 0.03\,M_{\odot}$.
This leads to relative large values of $P_{\rm spin}$
and $\dot{P}_{\rm spin}$ for the (mildly) recycled pulsar. 
As the mass-transfer rate is always highly super-Eddington
during the RLO, and the accreting neutron star will
be enshrouded by a thick (bloated) disk,
it is doubtful whether it will be observable as an {X}-ray
binary during this phase -- except very briefly just
at the onset and near the end of the RLO.

We followed the evolution of the donor star further on.
The donor continues to burn hydrogen in its light envelope.
At $t=186.4$ Myr (f) the helium burning is finally ignited 
($L_{\rm He}/L_{\rm H} > 10$)
in the core which now has a mass of $0.596\,M_{\odot}$.
After 70 Myr ($t=253.8$ Myr since the ZAMS) the core-helium
burning is exhausted (g). The $0.602\,M_{\odot}$ core then has a chemical
composition of 19\% {C}, 79\% {O} and 2\% {Ne}.
It is surrounded by a $0.016\,M_{\odot}$ envelope
(16\% {H}, 82\% {He}, 1\% {N}$_{14}$).
The central density is $\rho _{\rm c}=4.13\times 10^4$ g cm$^{-3}$
and $R_2=0.21\,R_{\odot}$. From here on the star contracts
and settles as a hot {CO} white dwarf.\\
We have now demonstrated a scenario for producing a BMSP with
the same characteristics as those listed in Table~1.

\subsection{The $P_{\rm orb}$--$M_{\rm WD}$ diagram}
In Fig.~2 we have plotted the calculated final orbital periods as a function of the
mass of the white dwarf companion (the remnant of the donor) for a given initial
mass of the donor star. 
The values for the BMSPs which originated from a binary with a low-mass companion
(the former LMXBs with $M_2 \la 1.8\,M_{\odot}$ located on the upper branch)
are taken from Tauris \& Savonije (1999). 
These white dwarfs are expected to be helium WD -- unless the initial orbital
period was very large ($P_{\rm orb} > 150$ days) so a relatively heavy
helium core developed prior to the RLO, in which case the helium core later
ignited forming a {CO} WD.\\
The final product of {X}-ray binaries with $M_2 > 2\,M_{\odot}$
are seen to deviate significantly from the low-mass branch. 
The reason is the former systems had a large mass ratio, $q\equiv M_2/M_{\rm NS}$ which caused
the binary separation to shrink initially upon mass transfer -- cf. Fig.~1. 
Such systems only ``survive''
the mass-transfer phase if the envelope of the donor is radiative or
slightly convective. 
This sets an upper limit on the initial orbital period for a given system.
If the donor is in a wide binary it develops a deep convective envelope prior to filling
its Roche-lobe and it will therefore expand rapidly
in response to mass loss which, in combination with the orbital shrinking,
will result in the formation of a CE and a (tidally unstable)
spiral-in evolution.\\
In Fig.~3 we show how the final orbital period and the mass of the WD depends 
on the initial orbital period for a binary with $M_2=4.0\,M_{\odot}$.
We notice that the question of initiating RLO before or after the termination
of hydrogen core burning (case A or early case B, respectively) is
important for these relations. 
For initial $P_{\rm orb} < 2.4$ days (case A RLO), $P_{\rm orb}^{\rm f}$
decreases with increasing $P_{\rm orb}$.
The reason is simply that in these systems the donor star is still on the
the main-sequence and the mass of its helium core, at the onset of RLO, increases
strongly with $P_{\rm orb}$ and therefore the amount of material to
be transferred (the donor's envelope) decreases with $P_{\rm orb}$.
Since the orbit widens efficiently near the end of the mass transfer,
when the mass ratio between donor and accretor has been inverted (cf. Fig.~1),
$P_{\rm orb}^{\rm f}$ will also decrease as a function of initial $P_{\rm orb}$.
However, for $P_{\rm orb}>2.4$ days
(early case B RLO) the final orbital period increases with initial
orbital period as expected -- the core mass of the donor only increases
slightly (due to hydrogen shell burning) as a function of initial $P_{\rm orb}$.

\subsection{The initial ($M_2,P_{\rm orb}$) parameter space}
In Fig.~4 we outline the results of our work in a diagram showing
the fate of a binary as a function of its initial $P_{\rm orb}$
and the value of $M_2$.
We conclude that {X}-ray binaries with $2\la M_2/M_{\odot} < 6$ 
can avoid a spiral-in and CE evolution
if $P_{\rm orb}$ is between 1--20 days, depending
on $M_2$.
If the initial $P_{\rm orb}$ is too short, the systems
will obviously enter a CE phase, since these
system always decrease their orbital separation when the mass transfer
is initiated\footnote{
In these narrow binaries the amount of available
orbital energy (a possible energy source for providing the outward
ejection of the envelope) is small and hence the neutron star
is most likely to spiral in toward the (unevolved) core of the donor,
forming a Thorne-Z\.{y}kow-like object.
In that case the neutron star will probably undergo
hypercritical accretion and collapse into a black hole.}.
On the other hand, if the initial $P_{\rm orb}$ is too large
the donor develops a deep convective envelope prior
to RLO and a runaway mass-transfer event is unavoidable,
also leading to a CE formation.
For systems with $M_2 \la 1.8\,M_{\odot}$ and initial $P_{\rm orb}<1$ day
the outcome is a BMSP with an ultra low-mass degenerate hydrogen star
(e.g. PSR~J2051--0827, cf. Ergma, Sarna \& Antipova 1998).

\section{Discussion}
We have now demonstrated how to form a BMSP with a relatively heavy
({He} or {CO}) WD companion without evolving
through a CE phase. 
If a substantial fraction of BMSPs have evolved through a phase
with super-Eddington mass transfer on a sub-thermal timescale (a few Myr),
this will eliminate the need for a long {X}-ray phase. This would
therefore help solving the birthrate problem between BMSPs and LMXBs
(Kulkarni \& Narayan 1988) for systems with $P_{\rm orb}^{\rm f} < 50$ days.\\
It has recently been suggested (Podsiadlowski \& Rappaport 1999;
King \& Ritter 1999) that Cygnus {X}-2 descended from
an intermediate-mass {X}-ray binary via a scenario which resembles
the one described here. We confirm that Cygnus~{X}-2 is a progenitor candidate
for a BMSP with a heavy WD.

It is seen from Fig.~2 that we can not reproduce the systems
with very massive {O-Ne-Mg} WD or the short orbital
periods ($\la 3$ days) observed in some systems with a {CO} WD. 
We therefore conclude that these binaries most likely evolved through
a CE phase where frictional torques were responsible for 
their present short $P_{\rm orb}^{\rm f}$ (cf. thin lines in Fig.~2 and 
gray area in Fig.~4).
These systems therefore seem to 
originate from binaries which initially had a relatively {\em large}
$P_{\rm orb}$ and case C RLO  -- otherwise if $P_{\rm orb}$ was small
the stellar components 
would have coalesced either in the spiral-in process, or as a result of
gravitational wave radiation shortly thereafter
(typically within 1 Gyr for systems surviving case B RLO and spiral-in). 

\acknowledgments
We thank the Parkes Multibeam Survey team
and the Swinburne Pulsar Group  
for releasing binary parameters prior to publication.
We appreciate comments from Ene Ergma on the issue of
forming BHWD systems.
T.M.T. would like to thank Gerry Brown for 
discussions and hospitality at Stony Brook.
T.M.T. acknowledges the receipt of a Marie Curie Research Grant
from the European Commission.

\clearpage
\begin{deluxetable}{llllcc}
\tablecolumns{6}
\tablewidth{0pt}
\tablecaption{Observed Pulsars with a ``heavy'' WD Companion}
\tablehead{
\colhead{}  PSR$\qquad$ & $P_{\rm orb}$  & f & $M_{\rm WD}^{\rm obs}$ & $P_{\rm spin}$ & $\dot{P}_{\rm spin}$\\
& (days) & ($M_{\odot}$)  & ($M_{\odot}$) & (ms) & }
\startdata
 J1904+04\tablenotemark{*}  & 15.75 & 0.0046   & 0.27  &  71.1 & \nodata \\
 J1810--2005\tablenotemark{*} & 15.01 & 0.0085   & 0.34  &  32.8 & $1.3\times 10^{-19}$\\
 J1453--58\tablenotemark{*}   & 12.42 & 0.13     & 1.07  &  45.3 & \nodata \\
 J0621+1002  & 8.319 & 0.0271   & 0.540 &  28.9 & $<8\times 10^{-20}$\\
 J1022+1001  & 7.805 & 0.0833   & 0.872 &  16.5 & $4.2\times 10^{-20}$\\
 J2145--0750 & 6.839 & 0.0242   & 0.515 &  16.1 & $2.9\times 10^{-20}$\\
 J1603--7202 & 6.309 & 0.00881  & 0.346 &  14.8 & $1.4\times 10^{-20}$\\
 J1157--5112\tablenotemark{**} & 3.507  & 0.2546 & $>$1.20& 43.6 & $<9\times 10^{-19}$\\
\hline
 J1232--6501\tablenotemark{*} & 1.863 & 0.0014   & 0.175 & 88.3  & $1.0\times 10^{-18}$\\
 J1435--60\tablenotemark{*}   & 1.355 & 0.14     & 1.10  & 9.35  & \nodata \\
 B0655+64    & 1.029 & 0.0714   & 0.814 & 196   & $6.9\times 10^{-19}$ \\
 J1756--5322\tablenotemark{**} & 0.453 & 0.0475 & 0.683   & 8.87 & \nodata \\
\hline
\enddata
\tablenotetext{*}{New pulsar, Parkes Multibeam Survey 
                  (Manchester~et~al. 1999).}
\tablenotetext{**}{New pulsar, Edwards et al. (1999).} 
\tablecomments{$M_{\rm WD}^{\rm obs}$ is calculated assuming
                 $M_{\rm NS}=1.4\,M_{\odot}$ and $i=60\deg$.}
\end{deluxetable}

\clearpage

\figcaption[]{The evolution of an {X}-ray binary with
$M_2=4.0\,M_{\odot}$ and $P_{\rm orb}=4.0$ days.
The left panel shows the evolution of $P_{\rm orb}$ as
a function of $M_2$ (time is increasing to the right).
The central panel gives the mass-loss
rate of the donor as a function of its age since the ZAMS.
The right panel shows the evolution of the mass-losing donor
(solid line) in an HR-diagram. The dotted line represents
the evolutionary track of a single $4.0\,M_{\odot}$ star.
The letters in the different panels correspond to one another
at a given evolutionary epoch -- see text for further explanation.}

\figcaption[]{The final $P_{\rm orb}$ as a function of
WD mass for different BMSPs. Next to each
curve is given the initial mass of the donor star (the
progenitor of the WD) used in our evolutionary calculations.
The free parameter in each curve is the initial $P_{\rm orb}$
(at the onset of the RLO). The curves in gray color represent
the formaion of BMSPs with helium WD,
while the black curves are BMSPs with {CO} WD.
The open circles on some of the curves indicate the
transition from case A to early case B RLO mass transfer
(i.e. whether or not the donor burned hydrogen in
the core at the onset of the RLO, Kippenhahn \& Weigert 1990).
The thin lines show the calculated parameters for systems which evolved
through a CE and spiral-in phase scenario assuming an efficiency parameter
of $\eta _{\rm CE}\lambda=1.0$
(e.g. van~den~Heuvel 1994).
The 12 observed BMSPs with a ``heavy'' WD companion
are marked with a star, see Table~1.}

\figcaption[]{The dependence of final orbital period (top) and 
mass of the WD (bottom),
on the initial orbital period, $P_{\rm orb}$.}

\figcaption[]{This plot illustrates the allowed parameter space
(white area) for producing BMSPs without evolving through a CE phase.
If $M_2 > 1.8\,M_{\odot}$ and the donor has
a deep convective envelope at the onset of mass transfer
(i.e. $P_{\rm orb}$ is large)
the system will evolve into a CE and spiral-in phase.
This is also the case if the initial period is very short
and $M_2 > 1.8\,M_{\odot}$. In the latter case the neutron star may 
collapse into a black hole.
}

\clearpage
\addtocounter{figure}{-4}
  \begin{figure*}[t]
    \psfig{file=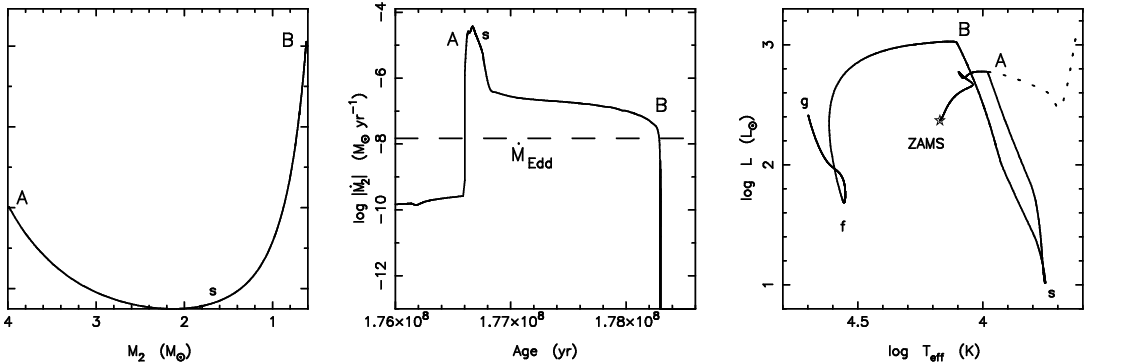,width=18.0cm}
     \caption{}
  \end{figure*}

\clearpage
  \begin{figure*}[t]
    \psfig{file=figL2.ps,width=18.0cm,angle=-90}
     \caption{}
  \end{figure*}

\clearpage
  \begin{figure*}[t]
    \psfig{file=figL3.ps,width=9.0cm,angle=-90}
     \caption{}
  \end{figure*}

\clearpage
  \begin{figure*}[t]
    \psfig{file=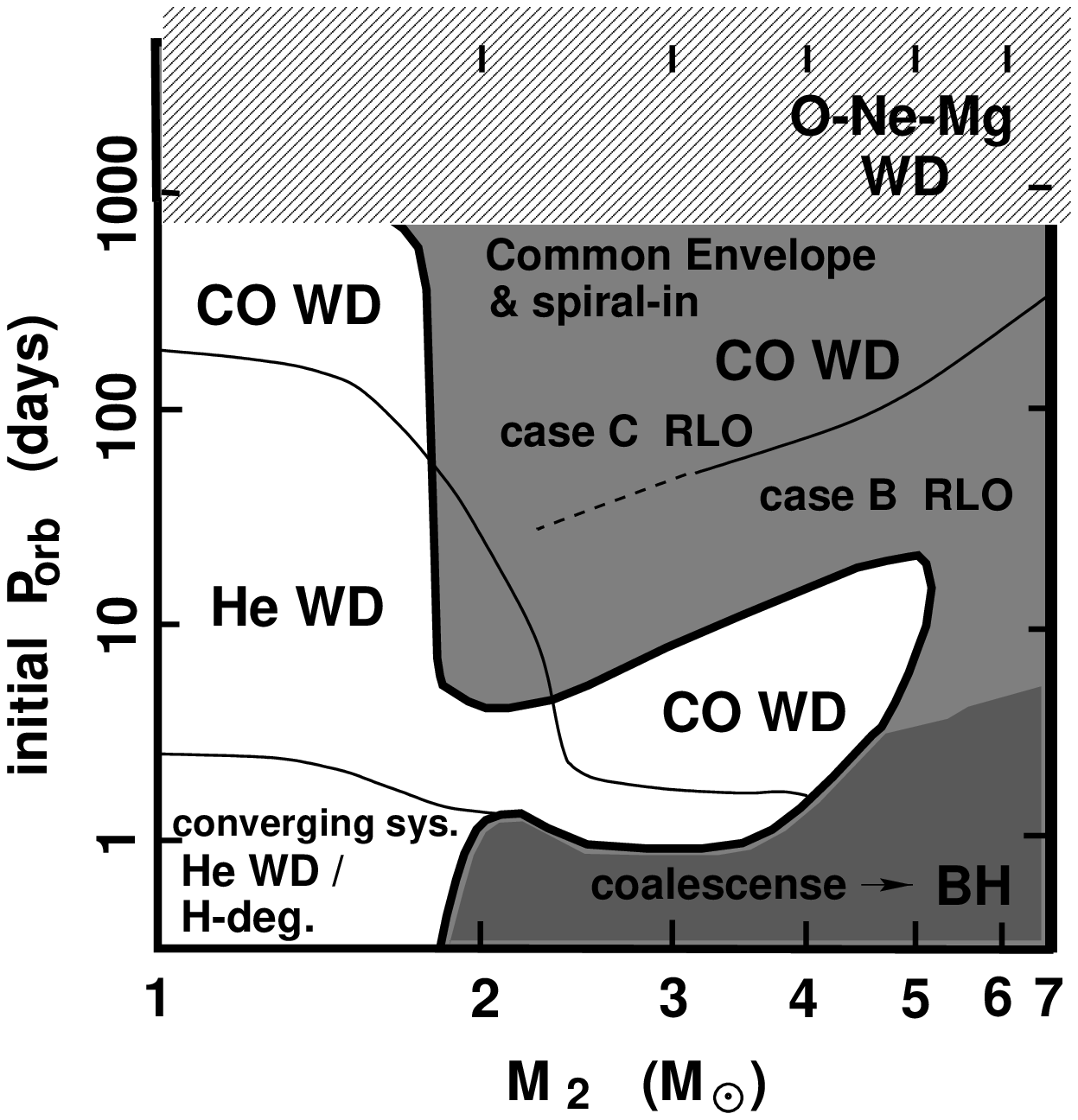}
     \caption{}
  \end{figure*}


\begin{references}
\reference{}Blandford R.D. \& Begelman M.C., 1999, MNRAS 303, L1
\reference{}Brown G.E., 1995, ApJ. 440, 270
\reference{}Chevalier R.A., 1993, ApJ. 411, L33
\reference{}Edwards R., et al., 1999, in preparation
\reference{}Ergma E., Sarna M.J. \& Antipova, J., 1998, MNRAS 300, 352
\reference{}Kalogera V. \& Webbink R.F., 1996, ApJ. 458, 301
\reference{}Kippenhahn R. \& Weigert A., 1990, Stellar Structure
            and Evolution, A\&A Library, Springer-Verlag
\reference{}King A.R. \& Ritter H., 1999, MNRAS, 309, 253
\reference{}Kulkarni S.R. \& Narayan R., 1988, ApJ. 335, 755
\reference{}Manchester R.N., et al., 1999, 
            to appear in: IAU Colloq. 177, Pulsar Astronomy -- 2000 and Beyond, 
            eds: M. Kramer et al. (ASP Conf. Series). 
\reference{}Narayan R. \& Yi I., 1995, ApJ. 444, 231
\reference{}Paczynski B., 1976,
            in: Structure and Evolution in Close Binary Systems,
            eds: \ P.P. Eggleton, S. Mitton, J. Whealan,
            Proc. IAU Symp.73, Dordreicht, Reidel, p.\ 75
\reference{}Podsiadlowski P. \& Rappaport S., 1999, ApJ. submitted,
            astro-ph/9906045
\reference{}Pols O.R., Schr\"{o}der K.P., Hurley J.R., Tout C.A.
            \& Eggleton P.P., 1998, MNRAS 298, 525
\reference{}Soberman G.E., Phinney E.S. \& van den Heuvel E.P.J., 1997,
            A\&A 327, 620
\reference{}Tauris T.M. \& Savonije G.J., 1999, A\&A 350, 928
\reference{}van den Heuvel E.P.J., 1975, ApJ. 198, L109
\reference{}van den Heuvel E.P.J., 1994, A\&A 291, L39

\end{references}
\end{document}